# Resilience in the Cyber World:
# Definitions, Features and Models

Elisabeth Vogel[1], Zoya Dyka[1], Dan Klann[1] and Peter Langendörfer[1,2]

[1]*IHP - Leibniz-Institut für innovative Mikroelektronik, Frankfurt (Oder), Germany*
[2]*BTU Cottbus-Senftenberg, Cottbus, Germany*
*{vogel, dyka, klann, langendoerfer}@ihp-microelectronics.com*

**Abstract**

Resilience is a feature that is gaining more and more attention in computer science and computer engineering. However, the definition of resilience for the cyber landscape, especially embedded systems, is not yet clear. This paper discusses definitions of different authors, years and different application areas the field of computer science/computer engineering. We identify the core statements that are more or less common to the majority of the definitions and based on this we give a holistic definition using attributes for (cyber-) resilience. In order to pave a way towards resilience-engineering we discuss a theoretical model of the life cycle of a (cyber-) resilient system that consists of key actions presented in the literature. We adapt this model for embedded (cyber-) resilient systems.

## 1. Introduction

The cyber landscape of the $21^{st}$ century is constantly growing and becoming increasingly complex covering areas such as telemedicine, autonomous driving etc. Our societies as well as individuals are highly dependent on these systems working correctly and 24/7. In order to be able to cope with the increasing complexity and the unprecedented importance of cyber systems, new and innovative methods and technologies have to be applied. The concept of resilience is getting increasing attention in this respect, which is reflected above all by the steadily growing number of publications on the topic. Figure 1 shows how the number of publications has increased since 2005. The diagram in Figure 1 shows only publications with the keyword *Cyber-Resilience*. Beneath its attention in science the concept of resilience already reached industry. US-American streaming provider Netflix is considered a pioneer in the application of resilience in the form of highly redundant infrastructure. But the principles of resilience are not only found in the hardware components of Netflix. The software architecture also demonstrates the application of various methods to increase resilience. The example of Netflix shows how important resilience is with increasing complexity [1].

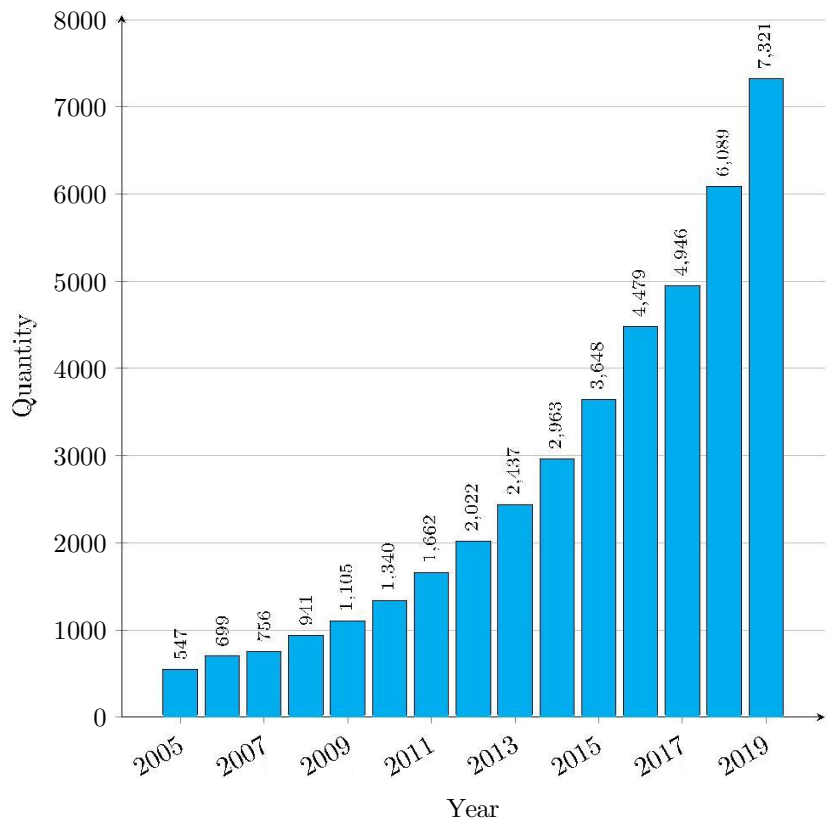

Figure 1: Number of publications with the keyword Cyber-Resilience from 2005 to 2019 [2].

But the term "resilience" is used in many ways in IT. In some cases, resilience is described as "extreme reliability" [3] or used as a synonym for fault tolerance [4], [5]. In [6] it is described that resilience is fault tolerance with the key attribute robustness. Anderson in [6] extended the definition of fault tolerance by the property robustness and called the new definition resilience. In recent publications, however, resilience is defined several times as an independent term [7] [8].

This clearly shows that the definition of resilience in IT as well as in other areas is certainly extensive and varied. While we admit that the term resilience is difficult to grasp we are also convinced that as a central concept of upcoming new IT-systems it needs a clear commonly accepted definition, metrics, etc. In order to achieve this, we give an overview of the definitions and properties of resilience and resilient systems in IT. For the purpose of a consistent presentation, we selected the publications as representative as possible. Table 1 shows the publications considered in this publication. This list of publications is of course only a very small selection, this is for a better overview. However, it also shows the different approaches of the authors, when defining resilience. Further literature with similar perspectives is also noted at the appropriate places. From the publications given in Table 1, the following information on resilience was extracted (where available): definitions, attributes, models. Furthermore, we discuss our model, which describes the structure of a resilient system. With the help of this model the development of a resilient system should be facilitated. These approaches are critically analysed and contrasted with our own holistic understanding of resilience. Finally, an example is used to show how approaches of resilience are already being implemented and in which direction the development could go in the future.

Table 1: The columns title, author(s), year show the papers we used as the main sources for this work, the column other relevant sources indicate additional publications that use the term resilience in a similar way as the main source in the same row.

| Year | Title & Author(s) | Further Sources |
|------|-------------------|-----------------|
| 1976 | **A Principle for Resilient Sharing of Distributed Resources [3]**<br>Peter A. Alsberg; John D. Day | |
| 2007 | **Release it! Design and Deploy Production-Ready Software [9]**<br>Michael T. Nygard | [10] |
| 2008 | **From Dependability to Resilience [11]**<br>Jean-Claude Laprie | [5], [12], [13] |
| 2011 | **Prologue: The scope of resilience engineering [14]**<br>Erik Hollnagel | [15], [16] |
| 2013 | **On the Constituent Attributes of Software and Organisational Resilience [17]**<br>Vincenzo De Florio | [4], [18] |
| 2015 | **Quantifying coastal system resilience for the US Army Corps of Engineers [7]**<br>Julie Dean Rosati; Katherine F. Touzinsky;<br>W. Jeff Lillycrop | [19], [20], [21], [22] |
| 2016 | **What's the Difference between Reliability and Resilience? [23]**<br>Aaron Clark-Ginsberg | [24], [25] |
| 2018 | **Systems Security Engineering: Cyber Resiliency Considerations for the Engineering of Trustworthy Secure Systems (NIST Special Publication 800-160, Volume 2) [8]**<br>Ron Ross; Richard Graubard; Deborah J. Bodeau;<br>Rosalie McQuaid | [26], [27], [28], [29] |

Furthermore, we discuss our model, which describes the structure of a resilient system. With the help of this model the development of a resilient system should be facilitated.

The rest of the paper is structured as follows. Section 2 shows different definitions of different authors, years and application areas for resilience. In section 3 the same authors as in section 2 are considered again, but this time under the aspect of attributes describing resilience. Section 4 describes the model of key actions, which was already briefly mentioned in section 2. This model of key actions will be reinterpreted in section 4 according to our ideas.

In section 5 we show current applications of resilience and the transferability to embedded systems. In addition, we discuss how to model and implement resilience.

## 2. Definitions

In the literature of recent years, there are many definitions of resilience, some of which differ considerably. As described in [4], the content of the definitions strongly depends on the respective fields of application. Resilience is derived from the Latin *resilire* and can be translated as "bouncing back" or "bouncing off". In essence, the term is used to describe a particular form of resistance.

How the term resilience is used in different disciplines (material science, engineering, psychology, ecology) is described in [30]. Also in computer science the term resilience has been defined several times from different points of view. As described in [4] for example, resilience is often used as a synonym for fault tolerance. However, recent publications show that this approach has been replaced by the view that resilience is much more than error tolerance (see [8]).

One of the first definitions was presented in [3] and describes the concept of resilience as follows:

*"He (remark: the user) should be able to assume that the system will make a "best-effort" to continue service in the event that perfect service*

70

*cannot be supported; and that the system will not fall apart when he does something he is not supposed to."*

In addition, [3] mentions attributes that constitute resilience as part of its definition. The attributes are the following: **error detection**, **reliability**, **development capability** and **protection against misuse** in the sense that the misuse of a system by individual users has only negligble effects on other users. According to Alsberg [3], these four attributes of a resilient system can be summarized as the attempt to describe extreme reliability and serviceability. In summary, a partial failure of a system should not have any effect on an individual user, so the system can be assumed to be highly reliable. Should nevertheless a partial failure or a defect occur, the best possible continuation of the services provided should be guaranteed. In extreme cases, this continuation can also be achieved by performing graceful degradation of services.

The approach of continuing a service of a system even under transient effects, permanent load or failures is also described in [9]:

*"A resilient system keeps processing transactions, even when there are transient impulses, persistent stresses, or component failures disrupting normal processing. This is what most people mean when they just say stability. It's not just that your individual servers or applications stay up and running but rather that the user can still get work done."*

According to Nygard [9], a system must remain stable in case of tensions or stress situations or failures. As consequence, involved (sub-) systems or possibly also users can still continue their work. The system must also be able to continue fulfilling at least its rudimentary functions despite any restrictions that may occur. The scope of these rudimentary functions may have been defined as part of the Risk Management, for example. Risk management also shows at what level of functional loss the entire system can no longer function according to its specifications.

In [11] resilience is described as the persistence of service delivery when changes occur that have system-wide effects. These changes can be functional, environmental or technological. In addition, changes can either be planned (for example: initialized by an update), the timing of their occurrence can be unpredictable, or they can be completely unexpected. The duration of changes is also taken into account: short-term, medium-term, long-term. This refers to the duration of the impact of the change on the system or a subsystem. Figure 2 shows the classification of changes schematically.

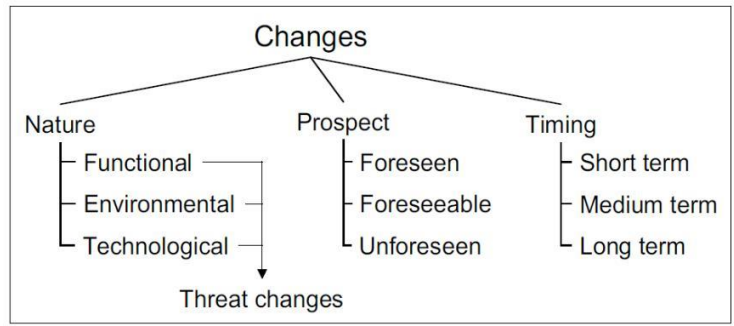

Figure 2: Change classification (source: [11]).

The paper by Laprie [11] proposes two definitions of resilience. The first definition is as follows:

> "The persistence of service delivery that can justifiably be trusted, when facing changes."

According to Laprie, this definition corresponds in principle to the original definition of reliability. In a second definition, Laprie offers an alternative which provides a more detailed description:

> "The persistence of the avoidance of failures that are unacceptably frequent to severe, when facing changes."

In [14], resilience is described as the ability to react to an event. This capability includes continuous observation of the performance of the system or a service provided by the system (self-monitoring), the recognition of future threats and learning from past failures.

In the collection of papers from 1985 [6], robustness was already mentioned in connection with resilience. About 30 years later, in [17] this connection is concretized. Resilience is defined as the trustworthiness of a software system to adapt to adverse conditions. The software system should accept and tolerate the consequences of failures, attacks and changes inside and outside the system boundaries. This is defined as an approach for robustness:

> *"Software resilience refers to the robustness of the software infrastructure and may be defined as the trustworthiness of a software system to adapt itself so as to absorb and tolerate the consequences of failures, attacks, and changes within and without the system boundaries."*

The definition of resilience was further specified in [17]. Florio [17] refers to the definition already mentioned in [4] and another definition in [31]. This definition states that resilience can be characterized as a measure of the persistence of both functional and non-functional features of a system under certain and unpredictable disturbances. After analyzing these two definitions, according to Florio, resilience is the ability to act and balance between two main behaviors:

1) Continuous readjustment with the aim of improving the system environment fit and compensating for both foreseeable and unforeseeable changes in the system environment.

2) Ensure that the said changes and adjustments from 1) do not affect the identity of the system. This means that its specific and distinctive functional and non-functional features should not be affected.

[7] deals with the management of water resources and was written from the perspective of the USACE[1]. According to Rosati [7], resilience is a cycle consist of anticipation, resistance, recovery and adaptation. Anticipation is the starting

---

[1]US Army Corps of Engineers

point of the cycle, while adaptation marks the end. The cycle is started by the occurrence of an event that affects the system in some way. This event is called a disruption. Specifically, Rosati defines resilience (in this case coastal resilience) as follows:

> *"(Coastal) resilience is defined as the ability of a system to prepare, resist, recover, and adapt to disturbances in order to achieve successful functioning through time."*

A disturbance occurs here as an effect of a hazard on the infrastructure, system, etc. A hazard is an environmental or adverse anthropogenic condition.

The article by Clark-Ginsberg published in 2016 [23] defines the ability of system to reduce the extent and the duration of disruptions as resilience. Disruptive events are not always predictable, but when they occur they are supposed to lead to a learning and adaptation effect of the system. Adaptation is crucial when it comes to realizing resilience against cyber accidents, since the cyber landscape is developing very rapidly. Clark-Ginsberg says in his article that errors must be detected and understood. It must be possible for the system to adapt to the errors or the error situation and a fast recovery must be guaranteed. The system must recover quickly after the occurrence of an error. If this is not possible, the error and the resulting faulty system environment must be dealt with appropriately.

The U.S. National Institute of Standards and Technology (NIST) [8] coins the term cyber-resilience to clearly distinguish its approach from the general definitions of resilience. Cyber-resilience is the following property:

> *"Cyber Resilience is defined in this publication as "the ability to anticipate, withstand, recover from, and adapt to adverse conditions, stresses, attacks, or compromises on systems that include cyber resources."*

According to NIST, the definition of cyber-resilience refers specifically to all entities that contain cyber-resources. A cyber-resource is an information resource that creates, stores, processes, manages, transmits, or disposes information in electronic form and that can be accessed over a network or by network methods. The definition of NIST can therefore be applied to a system, a mechanism, a component or a system element, a common service, an infrastructure or a system of systems, etc.

The publications selected here show that the type and scope of the definitions of resilience depend very much on the respective (informatics) application area. However, some key actions can be filtered out, which appear at least partially in all the publications considered here: Anticipating, resisting, recovering, adapting (to threats of any kind). In some publications, such as [7], [8], [32] these key actions are even mentioned explicitly. Table 2 shows which key action is mentioned in which of the publications considered here.

Table 2: Sources mentioning key actions

| No. | Publication | Meta-aspect | | | |
|---|---|---|---|---|---|
| **1** | Alsberg, 1976 |  |  |  | |
| **2** | Nygard, 2007 | |  | | |
| **3** | Laprie, 2008 | |  | |  |
| **4** | Hollnagel, 2011 |  | | |  |
| **5** | Florio, 2013 | |  | | |
| **6** | Rosati, 2015 |  |  |  |  |
| **7** | Clark-Ginsberg, 2016 |  |  |  |  |
| **8** | NIST, 2018 |  |  |  |  |

Anticipation   Recovery

Resistance   Adaptation

These key actions and our understanding of their interaction with each other are described in more detail in section 4. Each of the four key actions can be assigned different attributes and behaviors. They are described in the following section.

### 3. Attributes

Section 2 introduced the key actions anticipation, resistance, recovery and adaptation. Each key action comprises several attribute[2]. These attributes can be derived directly from the definitions or were explicitly mentioned in the publications. Table 3 shows the publications discussed in detail here and the key actions as well as the related attributes mentioned or deduced, respectively.

[3] describes the four main attributes of a resilient service. First of all, a resilient service must be able to detect and correct errors. Further, the resilient service must be so robust and reliable that a user expects the service not to fail. If the service is capable of always detecting $n$ errors and recovering from those errors, the $(n+1)$ error is not catastrophic. This only applies under the condition that the system offers perfect detection and recovery of $n$ errors. The resilient service is therefore able to anticipate the $(n+1)^{th}$ error in such a way that its negative consequences for the service can be minimized. This corresponds to a simple definition of evolvability. As a fourth key attribute, Alsberg [3] cites the ability of a resilient service to tolerate abuse by a single user in such a way that this abuse has negligible impact on the other users of the service. Alsberg does not specify misuse, but if a malicious and intentional action is assumed, then this misuse protection corresponds to the security feature of availability. Alsberg summarizes the following attributes: robustness, reliability, evolvability and security.

---

[2]In the publications presented here, the terms attribute, feature and measure were used synonymously. For the sake of clarity, only the term attribute will be used in this article, representing Feature and Measure.

Table 3: Key actions and attributes from the publications

| Publication | Key actions | Attributes | |
|---|---|---|---|
| 1976 Alsberg [3] | Anticipation<br>Resistance<br>Recovery | Robustness<br>Reliability<br>Evolvability | Security |
| 2007 Nygard [9] | Resistance | Reliability | Stability |
| 2008 Laprie [11] | Resistance<br>Adaptation | Reliability<br>Stability<br>Evolvability | Diversity<br>Assessability |
| 2011 Hollnagel [14] | Anticipation<br>Adaptation | Reliability<br>Evolvability | |
| 2013 Florio [17] | Resistance<br>Adaptation | Reliability<br>Evolvability | Integrity |
| 2015 Rosati [7]<br>2016 Clark-Ginsberg [23]<br>2018 NIST [8] | Anticipation<br>Resistance<br>Recovery<br>Adaptation | Robustness<br>Reliability<br>Evolvability<br>Security<br>Safety | Diversity<br>Assessability<br>Integrity<br>Stability |

[9] claims that stability under all conditions is the most important property of a resilient system. According to Nygard [9], this stability is directly related to the reliability of a resilient system. This connection is obvious, since a system that is not stable cannot be reliable either.

This understanding of stability and reliability is also illustrated in [11]. Assessability is also an important property, because a resilient system must be able to validate the correctness or plausibility of sensor data, for example. Laprie [11] also mentions diversity as another important basic property. Diversity can be understood here as a basic idea of redundancy, because according to Laprie, diversity in a system (of hardware components, for example) should prevent the occurrence of single point of failure.

In [14] it is also described that reliability is a key feature of resilience. The ability to detect a fault before it occurs is also essential. However, this only refers to faults that can be anticipated on the basis of existing information. A resilient system must be able to minimise the negative effects of a disturbance by anticipating it. This is done by constantly updating information about the disturbances that have already occurred and treated. This process can be understood as the ability to evolve. According to [17], the following attributes are essential for a resilient system: reliability, evolvability and integrity. Reliability and evolvability are related to resilience, as described in the previous definitions. Integrity, according to Florio [17], means that a resilient system does not lose its intention after adaptation or application of changes regarding a failure. This refers mainly to its functional and non-functional characteristics.

In the publications [7], [8], [23], the abilities to anticipate, resist, recover and adapt are directly mentioned as the four basic attributes or, as in NIST, the four basic goals of resilience.

Castano [33] assigns a large number of different attributes to resilience. Figure 3 shows this in a schematic representation.

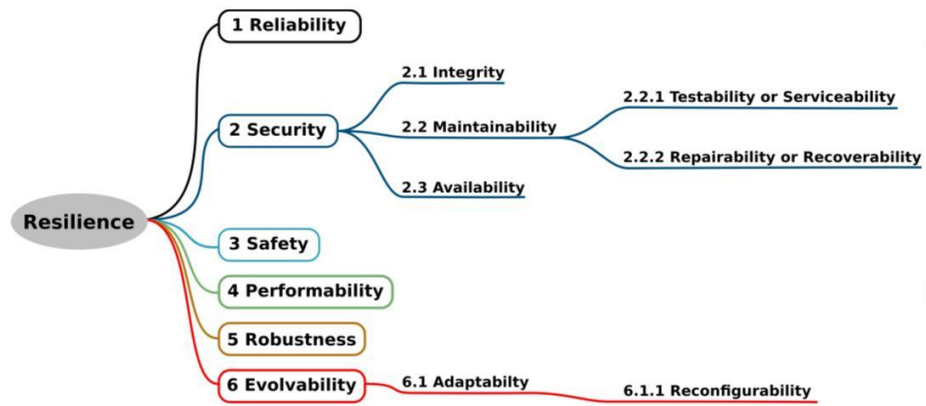

Figure 3: Attributes of resilience (image based on the following source: [33]).

Some of the attributes shown in Figure 3 and discussed in [33] build on each other or share identical approaches. An obvious example is the ability to evolve, which clearly lives from the ability to adapt. For an adaptation, in turn, some form of reconfiguration must always take place.

In the following section the key actions are described in more detail according to [7].

## 4. Model of key actions

The key actions of resilient systems were introduced in section 2. In this section their dynamic relationships are shown and explained.

The key actions are mentioned completely or partially in many publications but Rosati in [7] establishes a direct connection between them in the form of a cycle as shown in Figure 4. Rosati is not about implementing an IT system, but about the ability of different departments, components and participants to respond when a disaster occurs. This responsiveness must be constantly improved in the sense of Rosati's publication in order to keep (permanent) damage and loss of life as low as possible. The concept presented by Rosati is intended to implement a system-wide approach that will support the challenges of managing the United States' water resource infrastructure. However, this abstract concept can be transferred to IT systems. This will be discussed later in this section.

The cycle shown in Figure 4 is started when a disturbance occurs and is considered successful once it has been completed for a disturbance. A learning effect is considered to have occurred when there is a measurable improvement if the same fault occurs again in the future.

A disturbance is considered to be an adversity that has negative effects on the system. Negative effects are consequences that interfere with the intention of the system, i.e. in a certain way with its task. In the worst case, the compromised system can no longer fulfill its tasks. Disturbances can be caused by malicious, non-malicious, anthropogenic [3], non-anthropogenic, internal and external influences.

---

[3]influenced by humans, caused by

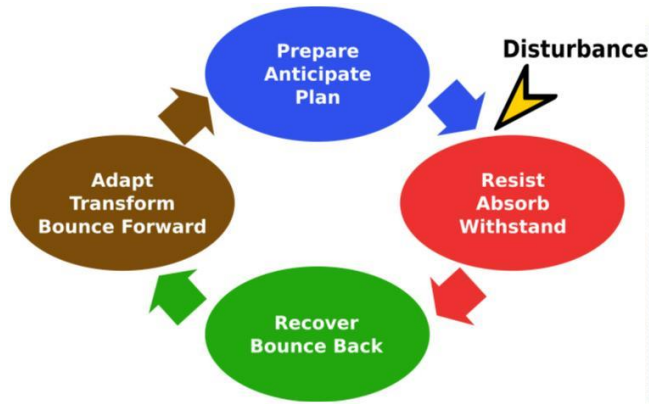

Figure 4: Presentation of key actions in a cycle (image based on the following source: [7]).

The four key actions shown in Figure 4 and the resulting cycle are to be understood according to Rosati as follows:

1. Prepare/Anticipate/Plan

   This key action includes a natural process or, under certain circumstances, an anthropogenic activity with the aim of preparing the system for a disturbance.

2. Resist/Absorb/Withstand

   This is the ability to withstand a disturbance while maintaining a certain level of functionality.

3. Recover/Bounce Back

   The lost functionality must be restored. If it is not possible to maintain functionality, the system shall be able to return to its original state.

4. Adapt/Transform/Bounce Forward

   The ability to adapt involves putting a system into a state that is better able to withstand or recover from the disruption. Ideally, this adaptation leads to reduced loss of functionality and a shorter recovery time. However, the process of adaptation only occurs when the cycle has been completed and applies only to this type of disturbance. Figure 5 shows the process of increasing resilience schematically.

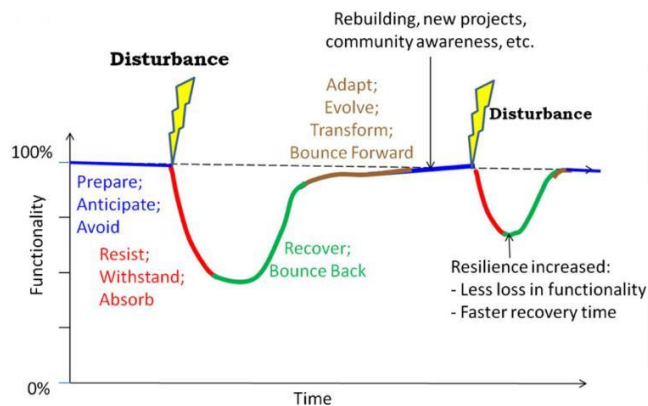

Figure 5: Schematic representation of the learning effect. A system is influenced by a disturbance and loses functionality. This loss must be resisted. Through the process of restoration and adaptation, a similar disruption will trigger a less severe loss of functionality in the future and the time to restore the system is shortened (source: [7]).

In the following discussion we present our idea of extending this concept of key actions to IT systems and critical infrastructure systems.

### Our model of key actions for Cyber-Resilience

The process of increasing resilience can be interpreted as the capacity for evolution. The system "learns" how to better deal with a disturbance that has already occurred at least once. This process is called evolvability.

The term "evolvability" originally comes from evolutionary biology and describes the ability of a living organism to bring about a change in its characteristics (attributes) by changing its genes with the aim of improving its (survival) abilities.

Two approaches are therefore important for evolvability, which build on each other:

1. Modification of genes
2. Changes in attributes resulting from 1.

The relation between genes and attributes can be transferred to a resilient system as follows: genes contain among other things the basic information for the evolution of the attributes of a living organism. The key actions represent the "genes" of resilience and the resilience itself has a multitude of attributes (reliability, safety, integrity, etc.). By modifying or improving the key actions (according to Figure 5) the properties of a resilient system can be improved.

The model of the four key actions from [7] is very well suited for the purpose described in [7]. It may also be applicable to an IT system, eventually in the area of critical infrastructure. For the IT area, however, we believe that modifications of the model are necessary. First, another essential key action must be added: **error analysis**. Furthermore, from our point of view, the key action resistance must be considered differently for a (distributed) IT system. An IT system must be able to continuously resist current and future disturbances. A disturbance does not necessarily have to be an event that immediately has negative consequences for the system. A disturbance can also be an attack with the aim of stealing information (e.g. private keys). Such actions, often carried out as side channel attacks [34], [35] do not cause any direct, immediately visible damage. However, if the extraction of the private key is successful, the system is considered broken from a security perspective and can no longer be described as resilient. This eliminates the key action resistance from the cycle. In our model resistance is divided into permanent and newly learned methods of resistance. This key action is continuously active, whereby "newly learned" methods of resistance are added to the permanent methods of resistance after the disturbance has been eliminated. Figure 6 shows our model of the key actions.

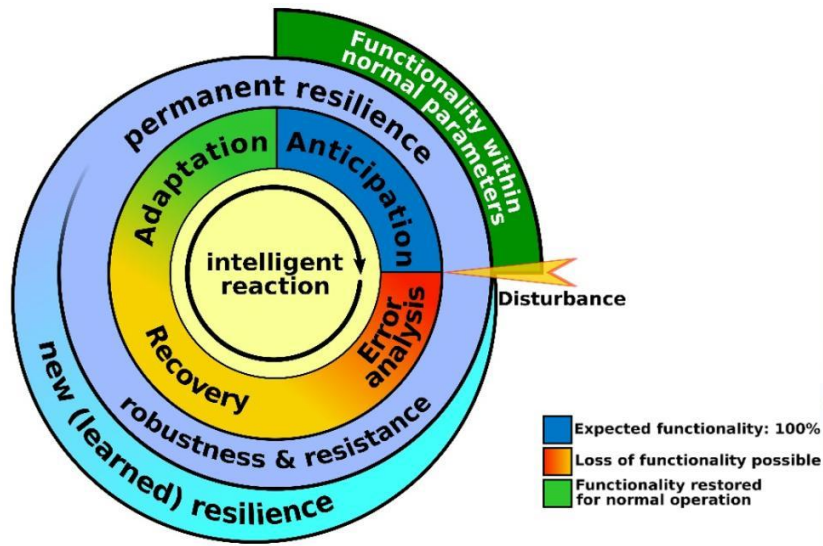

Figure 6: Schematic representation of the key actions anticipation, resilience (permanent methods and new (learned) methods), recovery and adaptation and the added key action error analysis (source: [36]).

The components of our model are described below. Please keep in mind that an IT system is under constant change i.e. its state concerning the key actions is also constantly changing. The boundaries between the key actions are not strict. There is hardly a rapid transition between the states. Thus, the individual methods of e.g. error analysis or recovery must also be considered. They interact with each other and partly also build on each other.

1. Anticipation

   Anticipation or the ability to anticipate is generally understood as the capability to predict future conditions, actions or events, taking into account only information already available. Anticipation is executed while the functionality of the system is within normal and expected parameters. The system should be able to anticipate possible disturbances from already known and/or collected information. This "previous knowledge" empowers the system to minimize the negative consequences of an anticipated disturbance in advance or even to prevent the disturbance completely.

Whether or not it is possible to prevent a disturbance naturally depends on the type of disturbance.

2. Error analysis

The error analysis basically covers four points: Error detection, error localization, error cause and error type determination. This key action is added to the four formerly used ones especially when disturbances occur which could not be anticipated in this form because no information was available (yet). In this way, error analysis correlates in a special way with anticipation. If a disturbance passes through the cycle of key actions several times, this disturbance can be better and better anticipated later on, which supports the process of evolvability considerably and shortens the time span for error analysis. This can be traced back to the stored information about the disturbance from previous runs. How good a disturbance can be anticipated depends on the type and complexity of the disturbance. For error analysis it is essential that the disturbance has been detected, understood, localized and possibly even predicted by the system. Additionally the cause of the disturbance can be helpful. However, the cause often cannot be identified or eliminated.

3. New (learned) methods of/for Resilience

This key action includes methods to resist, which are not permanently active. There is a pool of methods that are known to the system but inactive. If a disturbance occurs that is unknown to the system, i.e. could not be anticipated, additional methods to resist become active. If this is the case, the required method is added to the Permanent Resilience Methods (see point 4.) after the disturbance has been dealt with.

A corresponding algorithm decides on the basis of various parameters (type of disturbance, current operating environment, dangerousness of the disturbance, probability of the reoccurrence of the disturbance,...) whether and how long a method is added to the Permanent Resilience methods. It also decides when a method can become inactive again, e.g.

to save memory. Such an algorithm can be realized with methods of artificial intelligence.

4. Robustness & Resistance

Robustness & Resistance is the comprehensive key action that is always active. In general, this key action includes the ability to resist the negative consequences of a known disturbance. When a disturbance occurs, a loss of functionality must be expected. This loss of functionality as a negative consequence of a disturbance must be counteracted by methods of permanent resilience. This means, for example, the prevention of data loss or the forwarding of faulty data.

5. Recovery

During the recovery process, the disturbance that has occurred must be remedied as far as possible. It should be noted that the correction of a disturbance is not a process that necessarily only has to take place during recovery. Rather, the correction of a disturbance and the elimination of the negative effects of the disturbance is a continuous process that can also be started during the execution of resistance methods.

However, the process of eliminating a disturbance must be definitely completed with the completion of the recovery. Any lost functionality will, if possible, be reintegrated into the system according to the possibilities and mechanisms used for recovery. Irreparably damaged hardware, for example, can of course no longer be used, but a message could be sent to a maintenance team. This maintenance team can replace the hardware. It should be possible to replace hardware while the system is running.

6. Adaptation

According to Figure 5, Rosati [7] indicates that during the adaptation phase the system is still in a state in which full functionality has not yet been restored. We believe that for effective adaptation, the system should be in a state in which it is fully functional.

Adaptation essentially means that the system puts itself by self-modification in a new state, in that it can react more efficiently to this type of distur-
bance in the future. This means less loss of functionality and/or a shorter
recovery time.

However, the process of adaptation is by no means trivial. The system has
to take into account previous adaptations to other disturbances. These
must not be significantly negatively affected by the new phase of adap-
tation. Furthermore, the adaptation of the system should be checked
beforehand. This check can be made, for example, in a backup of the
system.

According to this modification of the model of key actions, the schematic rep-
resentation of the learning effect must be adapted. Figure 7 shows the new
schematic representation.

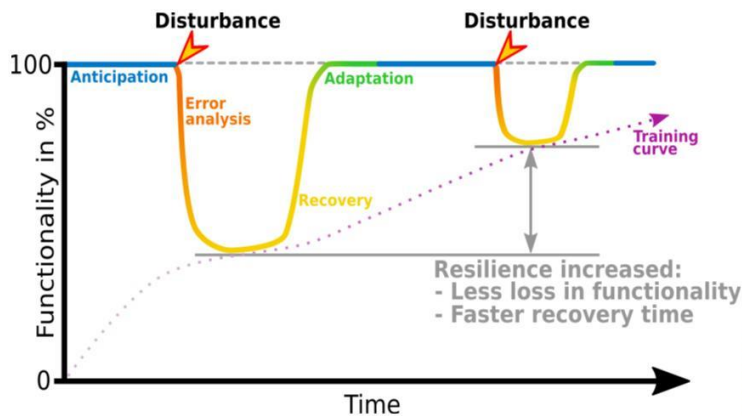

Figure 7: Schematic representation of the learning effect with the added key action error
analysis. The two key actions of resistance are to be regarded as decoupled and cover the
other 4 key actions of anticipation, error analysis, recovery and adaptation completely.

Based on the key actions and their relationships to each other, a resilient
system can be designed. For this purpose, in addition to the consideration of
the key actions, information from the following areas, among others, must be
taken into account:

- Mechanisms for error detection

- Mechanisms for attack detection (e.g. IDS [4])
- Mechanisms for error correction
- Mechanisms for fault localization
- Types of errors
- Mechanisms for pattern recognition

These mechanisms are to be applied according to the requirements of the system and the available options. Particular attention is required in the area of error detection. As a rule, errors are only detected after they have caused (negative) consequences or accidents. Thus it is obvious that the cycle of key actions consists of partially interlocking key actions that are directly or indirectly interdependent. For example, at the moment when an error has a negative impact on the system, the system must be able to resist this impact.

Thus, in order to achieve resilience Cyber-Physical System of Systems [5] needs to be capable to handle different adverse conditions, of which some might be expected while others are not.

CPS(oS) must be robust to specified disturbances. This includes specified manipulation and attacks. Specified disturbances are disturbances that could be expected during the development phase. This robustness can be realized e.g. with special materials, which guarantee the robustness of the design in the specified working area (the expected working conditions). These can be, for example, wires made of metal that have a higher melting point and can therefore meet the set requirements.

In addition, this robustness can be achieved by using redundancy. Redundancy is used according to the specified working conditions to ensure fault tolerance.

CPS(oS) must also be resistant to unspecified disturbances. Unspecified disturbances are disturbances that had to be expected but for which no reaction scenario was defined by the system. CPS(oS) must be able to resist unspecified disturbances (at least partially). It is essential that unspecified disturbances can be detected, minimized, predicted or even avoided at an early stage if they occur repeatedly. Unspecified disturbances can be triggered by the following causes, for example:

- Excessive deviation of the physical parameters of the environment from the specified working range

- Too frequent (even not strong) deviations of the physical parameters of the environment from the specified working range

- Too short reaction time

- (dynamic) changes in the specified working or analysis range (e.g. increased operating voltage)

- Individualized work or analysis areas

---

[4]Intrusion Detection System

[5]CPS(oS)

In short, we define resilience for CPS(oS) as follows:

> *A CPS(oS) is resilient if it has the ability to react to specified and unspecified disturbances in a way that preserves its function and reacts quickly. This reaction also includes the early detection, minimization, prediction or even avoidance of disturbances.*

According to our definition we discuss in the following four different types of systems i.e. primitive system without error handling (1), fault-tolerant system (2), resilient system (3) and total-resilient system (4) shown in Figure 8.

A system as the one shown in Figure 8.1 without any form of error detection should not be in use today, especially not in the area of critical infrastructure. This system is not able to deal with a fault, understand it or resist its negative effects. Figure 8.2 shows a fault-tolerant system. These systems are currently the ones most commonly used. If this system detects an error, this error can often be corrected. If the error cannot be corrected, the user is informed. If an error is not detected, e.g. because it has no direct effect on the functionality of the system, it is a corrupt system that can, among other things, supply incorrect data. Figure 8.3 shows our idea of a resilient system.

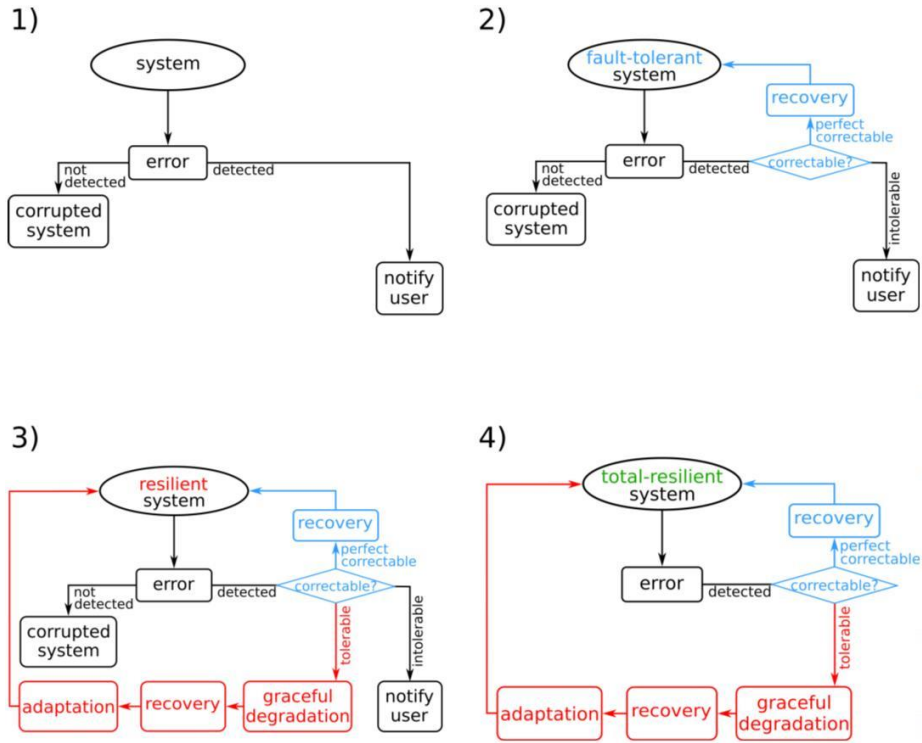

Figure 8: Schematic representation of four different systems: (1) primitive system without error handling, (2) fault-tolerant system, (3) resilient system, (4) total-resilient system.

This idea was developed using our model of key actions. The probability that an error was not detected (corrupt system) or that an error was detected but cannot be handled (inform users) is given but should be very low. Figure 8.4 shows a very extreme case of a resilient system: an absolute, i.e. totally resilient system. In a totally resilient system, every error is detected without exception and these errors can be handled. The system will also be able to make intelligent adjustments to an error, just like in a resilient system, so that it can react better to the error in the future.

In the current development process for highly reliable systems in critical infrastructure, it makes sense to develop increasingly resilient systems and to make fault-tolerant systems resilient. However, resilience is an extremely dynamic concept, as the many different definitions show. Ultimately, however, resilience is not only the ability to deal with errors, but also the realization that errors will definitely happen. It is irrelevant whether these errors are symptoms of a malicious attack, environmental influences or wear and tear. In order to effectively implement resilience, a holistic understanding of the system design and the threat situation is necessary. Also, such a system architecture must offer possibilities for expansion (during normal operation), since resilience is subject to constant change due to its dynamics.

## 5. Conclusion and Future Work

Cyber-Resilience encompasses more than security and reliability. Cyber-Resilience also deals with the ability of a system to make autonomous decisions according to the situation in which the system is located. These autonomous decisions can be modeled using artificial intelligence methods. However, these methods of artificial intelligence must also satisfy the conditions of security and reliability.

Currently Netflix is the most known example of a resilient system. Netflix's infrastructure and applications use a high degree of redundancy to implement the idea of Cyber-Resilience in the best possible way. In addition to redundancy, Netflix uses other mechanisms to ensure that high availability is maintained. These include extensive empirical checks of the resilience (the checks are performed live and during normal operation), rapid isolation of errors, the ability to quickly perform fallback, rollback and failover, and constant logging and monitoring of all activities in the system. All these mechanisms enable Netflix to guarantee almost continuous availability to users and to react to unexpected (negative) events [1].

Redundancy is a very powerful tool to achieve Cyber-Resilience, but it is not universally applicable. The level of redundancy used by Netflix to ensure Cyber-Resilience is unthinkable for embedded systems, for example. Redundancy requires a lot of physical space, especially in the area of hardware, but embedded systems cannot be extended at will. This means that one has to work with the given form factor and cost limitations. At the same time, however, it must also be determined what Cyber-Resilience means for an embedded system. The objectives are the same as those of Netflix, but the requirements are limited. The embedded system must also be capable of self-observation (logging and monitoring) and it must be able to predict and detect (negative) changes/events in order to react. The first big challenge is the (correct and complete) detection of a negative change, because if a system is not able to do so, it cannot take appropriate countermeasures.

If negative events were detected correctly, the system must react to these events. The way the system reacts to such an event is another major challenge. It is not sufficient to say that a system must always react to error X with countermeasure Y. Embedded systems are highly complex and components are sometimes highly interdependent. The system must react in a way that the negative event can be countered and at the same time own functionalities are not or only temporarily impaired. Also (sensor) data must not be corrupted or lost. In addition, various negative influences can occur almost simultaneously. The different countermeasures must not hinder or even prevent each other under any circumstances.

The third major challenge is the optimal selection of mechanisms that can be used preventively against negative events. Redundancy would be one such mechanism, but it is only of limited use in embedded systems. A further mechanism would be the isolation of different components of an embedded system, so that negative events from which errors/disruptions arise are limited to one component. In this way, cascading to other components can be prevented. Of course a complete isolation is not possible, but there are mechanisms that are summarized under the term loose coupling that help to keep the interrelationships as small as possible.

So there is a very wide range of different methods that can be used to achieve Cyber-Resilience. Especially for embedded systems it has to be planned exactly which methods should be used. The reasons for this are, among others, the limited space, the limited storage and computing capacity (also with regard to the use of software solutions for Cyber-Resilience) as well as the available financial means. In addition, the location and the degree of criticality of the system (critical infrastructure) play an important role.

The great challenge of Cyber-Resilience is the planning and development of systems that fulfill selected aspects of security and reliability. In addition, a system must be able to make intelligent decisions in order to defend itself efficiently against negative effects.

### Future Work

One approach would be to develop a design kit for cyber-resilient systems. To develop such a design kit it is first necessary to collect and classify different methods for achieving Cyber-Resilience. A classification only makes sense if the methods have been proven to have a positive effect on a system. Theoretical methods must first be tested.

The classification of the methods could then look as follows: effort of implementation, type and extent of effect on the system, possibilities of implementation, etc. The next step is to examine what dependencies exist between these methods. This refers to how these methods influence each other within a system. This allows negative dependencies to be taken into account. This analysis of the dependencies can be implemented e.g. with the help of artificial intelligence.

The idea would be a semi-automatic design kit that combines different methods for Cyber-Resilience and can select and combine them with the help of artificial intelligence in a way that is suitable for the system. The result is a theoretical model that can be used for the practical development of a cyber-resilient system.